# Adhesion-assisted nanoscale rotary locomotor in non-liquid environments

Jinsheng Lu[1], Qiang Li[1], Cheng-Wei Qiu[2] and Min Qiu[1*]

[1]State Key Laboratory of Modern Optical Instrumentation, College of Optical Science and Engineering, Zhejiang University, 310027, Hangzhou, China

[2]Department of Electrical and Computer Engineering, National University of Singapore, 4 Engineering Drive 3, Singapore 117583, Singapore

*minqiu@zju.edu.cn

Rotation in micro/nanoscale provides extensive applications in mechanical actuation[1,2], cargo delivery[3,4], and biomolecule manipulation[5,6]. Light can be used to induce a mechanical rotation remotely, instantly and precisely[7-13], where liquid throughout serves as a must-have enabler to suspend objects and remove impact of adhesion. Achieving light-driven motion in non-liquid environments faces formidable challenges, since micro-sized objects experience strong adhesion and intend to be stuck to contact surfaces. Adhesion force for a usual micron-sized object could reach a high value[14,15] (nN - μN) which is several orders of magnitude higher than both its gravity (~ pN) and typical value of optical force (~ pN) in experiments[16]. Here, in air and vacuum, we show counter-intuitive adhesion-assisted rotary locomotion of a micron-sized metal nanoplate with ~30 nm-thickness, revolving around a microfiber. This locomotor is powered by pulsed light guided into the fiber, as a coordinated consequence of photothermally induced surface acoustic wave on the nanoplate and favorable configuration of plate-fiber geometry. The locomotor crawls stepwise with sub-nanometer locomotion resolution actuated by designed light pulses. Furthermore,

**we can control the rotation velocity and step resolution by varying the repetition rate and pulse power, respectively. A light-actuated micromirror scanning with 0.001° resolution is then demonstrated based on this rotary locomotor. It unfolds unprecedented application potential for integrated micro-opto-electromechanical systems, outer-space all-optical precision mechanics and controls, laser scanning for miniature lidar systems, etc**.

Light-driven rotation has been well demonstrated through transferring linear[8, 9] or angular[10-13] momentum to micro-sized objects in liquid environment. It is easy to be taken for granted that these optical manipulation phenomena will be naturally present and held in non-liquid environments. However, dominant adhesion in non-liquid environments simply rules out such hallucination. Adhesion seriously impedes operation of rotary motors working in these way of momentum transferring, and thus liquid is widely used to erase the unwanted impact of adhesion. Distinguished from this long held captivation, we report a light-actuated rotary locomotor for which the adhesion counter-intuitively becomes the key enabler for the rotation with the synergetic assistance from surface acoustic wave (which is generated due to thermo-elastic expansion by heating of absorbed pulsed light) and geometrical configuration. The experimental setup is shown in Fig. 1a. A uniformly fine-drawn optical microfiber is suspended in air or in vacuum, where a gold nanoplate (NP) is placed on it by using a probe. This NP-microfiber system can be seen clearly in the scanning electron microscope (SEM) image (Fig. 1b). We firstly attempted to switch a CW laser on/off, a subtly weak azimuthal movement of a gold NP was spotted at the instantaneous ON/OFF moments. This movement is due to the expansion/contraction of the gold NP[17]. That accidental effect, which is highly inspirational, triggers us to deliver a pulsed supercontinuum light into the microfiber, thereby enabling almost stably continuous azimuthal rotation. The gold NP (i.e., the locomotor) starts to revolve around the microfiber once the light pulses are guided into the microfiber. The gold NP is tightly attached on the surface of the microfiber by adhesion force during rotation. The adhesion force between the NP and the microfiber is measured to be ~ 6 nN while the gravity of the NP is only several pN. We also conducted this experiment in liquid where the adhesion force is rather small. In this case, the gold NP leaves the microfiber and does not rotate around the microfiber anymore. This means that the adhesion force is indispensable for this rotation.

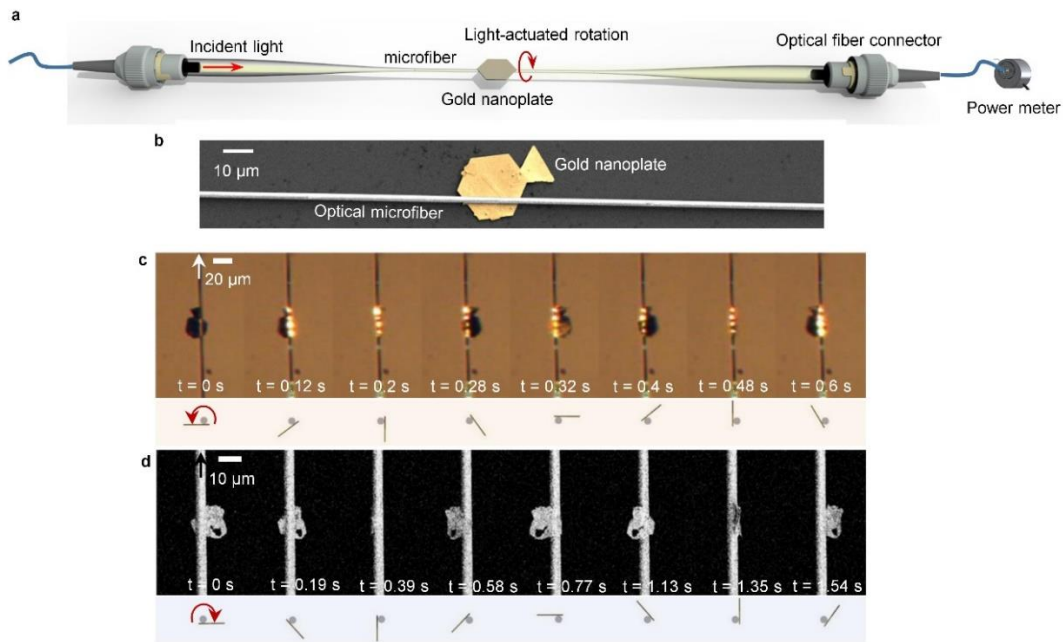

**Figure 1 | Light-actuated rotation of a locomotor in air and in vacuum. a,** Schematic of experimental configuration showing that pulsed supercontinuum light (pulse duration, 2.6 ns; repetition rate, 5 kHz; wavelength, 450 nm ~ 2,400 nm) is delivered into a microfiber and light power is measured by a power meter at output end. The microfiber is suspended in air or in vacuum, and a gold NP is placed on it and then rotates around it actuated by the pulsed light. **b,** False-color scanning electron micrograph of a gold NP (side length 11 μm; thickness 30 nm) below a microfiber with a radius of 880 nm. Note that the NP-microfiber system shown in **b** is placed on a silicon substrate after rotation experiments. **c,** Sequencing optical microscopy images of the anticlockwise revolving gold NP around the microfiber in air (Sample A, 5 kHz). The measured average light power is 0.6 mW. **d,** Sequencing SEM images of a clockwise revolving gold NP (long side length 10.5 μm; short side length 3.7 μm; thickness, 30 nm) around a microfiber (radius, 2 μm) in vacuum. The measured average light power is 1.5 mW. Arrows in **c** and **d** represent the direction of light propagation. Gray circles and yellow lines below **c** and **d** denote microfiber and NP, respectively. Red curve arrows indicate rotation direction of NP.

The optical microscope images (Fig. 1c) sequentially demonstrate the rotation of the locomotor in air. The locomotor rotates at ~1.4 Hz anticlockwise (viewed in the direction of the light propagation). The rotation direction is dependent on initial relative location of the NP on the microfiber. The NP rotates anticlockwise if the NP is initially adhered on the bottom of the microfiber and center of the NP is shifted to the left half of the microfiber, and otherwise it rotates clockwise (see Fig. 1d for comparison).

The locomotor can not only work in air but also in vacuum where the gas pressure is about 9 orders of magnitude lower than that in air environment. Therefore, the impacts due to the photophoretic force, as one kind of commonly light-induced force[18], can be ruled out in vacuum. Figure 1d shows a gold NP revolving around a microfiber in vacuum. The NP rotates at ~ 0.5 Hz clockwise.

The rotation speed of the locomotor is found linearly proportional to the repetition rate of light pulses, as shown in Fig. 2c. The rotation speed of the NP can be gained from the period of effective width variation of the NP recorded in frames of experimental videos (Fig. 2a). The power of every light pulse remains the same and the repetition rate of the emitting light pulses increases. In this case, the rotation speed increases linearly, showcasing that a single light pulse can actuate the locomotor to rotate an extremely fine angle and the rotation angle for every light pulse keeps almost constant, independent on the repetition rate. This is further confirmed by following experiments.

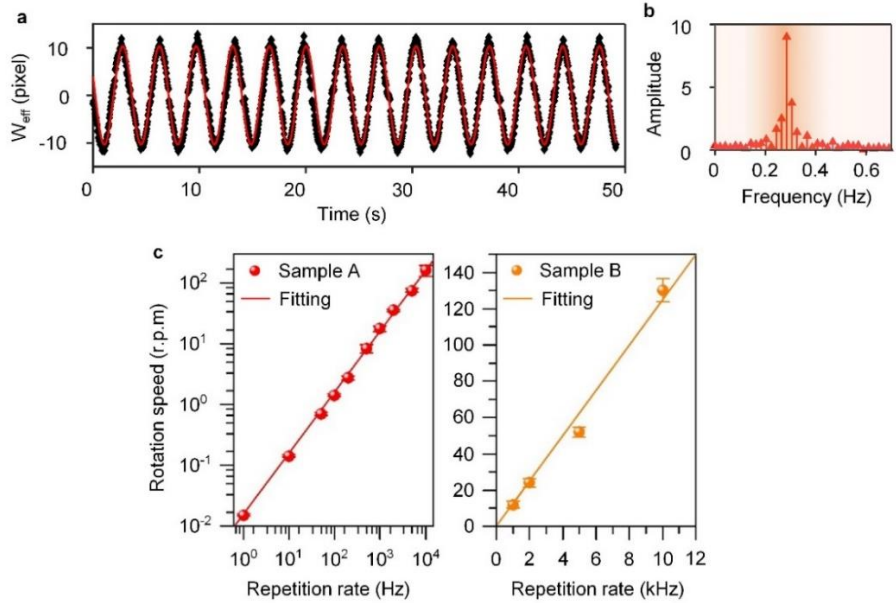

**Figure 2 | Relationship between rotation speed and repetition rate. a,** Effective width ($W_{eff}$) of the NP obtained from every frame of experimental videos (Sample A, 1 kHz). For convenient, the effective width of the NP is presented with pixel length. **b,** Fourier transformation of the effective width to obtain its variation frequency (i.e. rotation speed of the NP). **c,** Light-actuated rotation speed of the locomotor increases linearly with repetition rate of light pulses, and different samples give similar results. The power for every light pulse keeps the same when the repetition rate is changed. Error bars are the variance of rotation speed.

As mentioned above, every single light pulse can actuate the locomotor to rotate a constant angle. Based on this, a desired rotation angle of the locomotor can be achieved if a specific number of light pulses are sent to the microfiber. We used a waveform generator to produce triggering signal which can task the light source to emit a specific number of light pulses, as shown in Fig. 3a. The locomotor rotates step by step for intermittent light pulse burst. The step angle increases linearly with the light pulse number as shown in Fig. 3b. The locomotor rotates about 0.1° for every single light pulse. The step angle for every single light pulse can be smaller (e.g. 0.018°) if the power of single light pulse decreases. Stepping rotary motion of the locomotor actuated by the intermittent light pulse burst is shown in Fig. 3c. The angle between the NP and the microfiber is calculated using projection method.

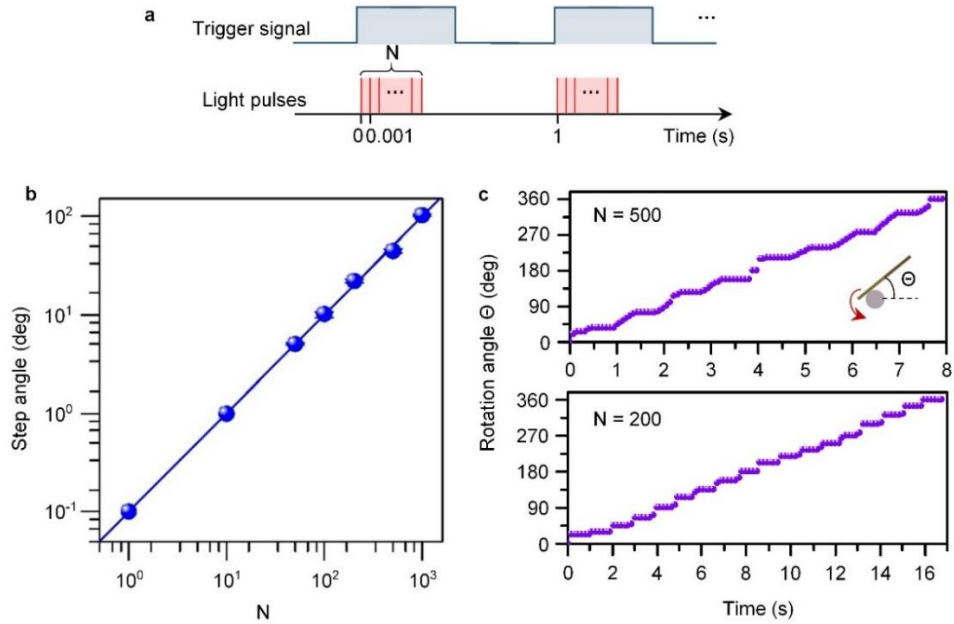

**Figure 3 | A stepping rotary locomotor. a**, Schematic showing that a specific number (N) of light pulses emit at 1 kHz repetition rate when light source senses a positive edge on every trigger input. The 1 Hz electric trigger signal is generated by a waveform generator. **b**, Step angle of the locomotor increasing linearly with the light pulses number (N) for one of trigger input. The locomotor rotates about 0.1° for every single light pulse. **c**, Stepping rotation of the locomotor when the light pulses numbers (N) are 500 and 200.

A pulsed light source with single wavelength (1,064 nm) can also be used to drive the locomotor to rotate. Continuous wave (CW) laser sources with different wavelength (e.g. 532 nm, 980 nm, and 1,064 nm) are used in experiments, but no rotation happens in these cases even when the laser power is high enough to melt the gold NP (thus optical force can be ruled out as the driving force for the rotation). This further indicates that pulses play an essential role in generating motion of the locomotor. Pulsed light can excite coherent phonons and induce lattice expansion and contraction during the propagation of acoustic wave[19]. For example, surface acoustic wave (i.e., Rayleigh wave) is generated on a metal surface when pulsed laser is focused as a line on the surface[20-22]. Pulsed light absorbed by the metal locally heats the metal, leading to local thermo-elastic expansion and generation of this surface acoustic wave. The progressive Rayleigh wave can produce a friction force to drive a slider placed on the surface to move[23-25]. Similarly, in this work, the pulsed-light-induced Rayleigh wave, which is generated on the gold NP and acts on microfiber surface, drives the NP to locomote on the surface of the microfiber.

  As shown in Fig. 4a, the pulsed light interacts with the gold NP in a line-shaped area through evanescent field in the vicinity of the microfiber surface, in analog with a pulsed laser focused as a line on the gold NP using cylindrical lens. Such Rayleigh wave interacts with contact surface and generates a friction force as a consequence. According to the characteristics of Rayleigh wave[26], the neighboring atoms at individual positions of the gold NP surface expand/contract, forming an elliptical motion in a collective fashion as Fig. 4b illustrates. The atoms have a maximum horizontal velocity component at ridges of the Rayleigh wave. This motion is conveyed to the microfiber surface through friction force, which drives the intended locomotion of the gold NP. Note that the ridges shown in Fig. 4b just simply represent the Rayleigh wave for convenience and the number of ridges generated in the gold NP should be less than one. It is because the velocity of the acoustic wave in gold film is[19] about 3,000 m/s and the frequency of nanosecond light pulse induced acoustic wave in metal film is[22] ~ MHz. The wavelength of the Rayleigh wave in gold NP should be several tens of micrometers to hundreds micrometers, which is larger than the size of the gold NP. The gold NP is slightly bent to make sure that a certain amount of area of the gold NP can be tightly attached to the microfiber surface due to the strong adhesion force. Therefore, the gold NP turns simultaneously when it crawls upon the curvilinear microfiber. Furthermore, the adhesion force enhances the interaction between the Rayleigh wave and microfiber surface just as the assisted role of "preload"[25]. As a result, the force which drives the plate to locomote is enlarged by the adhesion force.

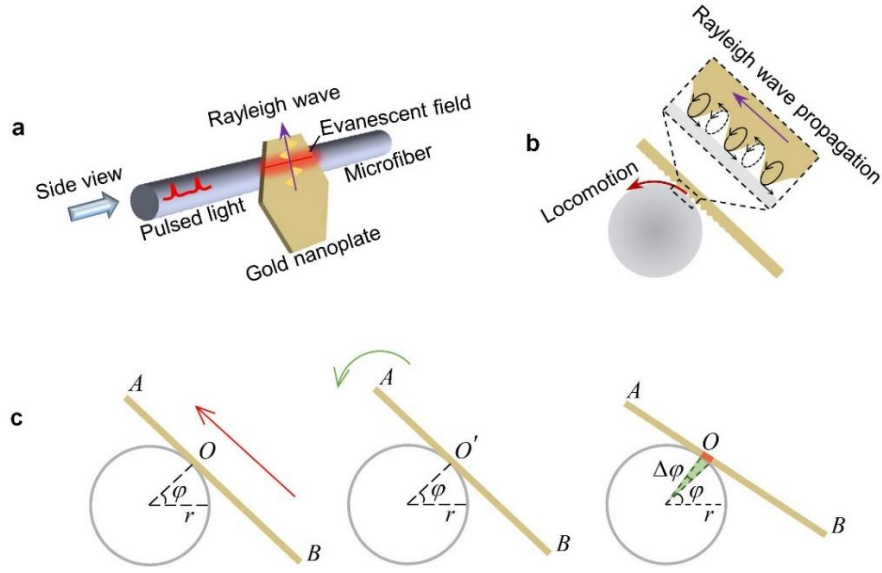

**Figure 4 | Surface-acoustic-wave-driven mechanism of the locomotor. a**, Schematic diagram showing a Rayleigh wave is generated on surface of a gold NP by line-shaped evanescent field outside a microfiber. **b**, Schematic diagram (side view) showing locomotion of the NP on microfiber surface driven by Rayleigh wave induced friction force. Black ellipses in partial enlargement of **b** represent collective motion pattern of surface gold atoms caused by Rayleigh wave propagation. **c**, Decomposition of movements showing synergic crawling and turning of the gold NP during one light pulse. Gray circle and yellow rectangle indicate microfiber surface and NP, respectively. *A, B, O (O')* in **c** represent the head, the end, and the contact point of the NP, respectively. Note that the contact point represents the center point of the contact area. The NP crawls upward (shown from left to middle in **c**) and the contact point of the NP changes from $O$ to $O'$ ($\frac{AO}{OB} < \frac{AO'}{O'B}$). The NP turns at a slight angle ($\Delta\varphi$) (shown from middle to right in **c**). The NP locomotes a distance ($\Delta\varphi \cdot r$) after this series movements and the contact point of the NP remains unchanged.

Figure 4c elaborates the decomposition of the movements of the gold NP during actuation of one light pulse. The surface-acoustic-wave-induced friction force drives the gold NP to crawl upward (from left schematic to middle schematic in Fig. 4c), and simultaneously the adhesion force causes the gold NP to turn (from middle schematic to right schematic in Fig. 4c). Resultantly, the relative contact area on the gold NP remains unchanged after a small rotation actuated by one pulse. When the NP is actuated by a series of pulses, this synergetic coordination between crawling and turning ensures that the gold NP can rotate around the microfiber continuously and will not be detached from the microfiber. The gold NP rotates a small angle $\Delta\varphi$, i.e., locomotes a distance $\Delta\varphi \cdot r$ after one pulse actuation. For one of experimental results, the rotation angle for one light pulse is 0.018° and the radius of the microfiber is 900 nm. Therefore, the locomotion distance for one light pulse (i.e., locomotion resolution) is 0.28 nm, which could largely correlated with vibration amplitude of induced Rayleigh wave.

Our rotary locomotor could be used as a micromirror to deflect light beam, which is important for integrated optomechanics and outer-space applications in non-liquid environments. Pulsed light is guided into a microfiber and used to drive a NP to rotate. A 532 nm CW laser beam is focused on and reflected by the NP. A screen in far field is used for displaying projected laser spot (Fig. 5). The NP rotates anticlockwise, and the rotation angle for one light pulse (i.e. scanning resolution of this micromirror) is only about 0.001°. The reflected light rotates with doubled speed subsequently, and we can see that the projected laser spot moves upward on the far field screen. The measured position of the laser spot on the screen is in good agreement with the theoretical expectation (Fig. 5c).

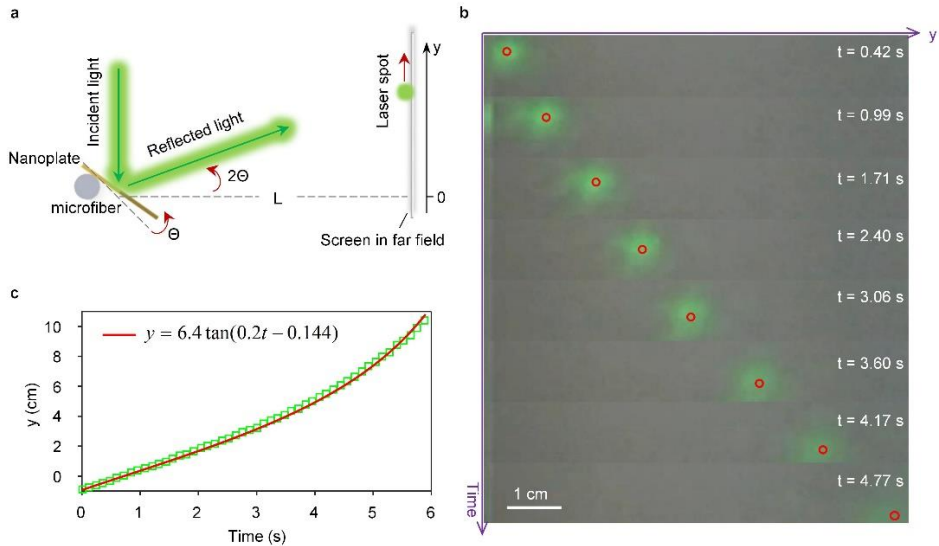

**Figure 5 | One example application, demonstrating a micromirror for laser scanning. a,** Schematic representation of a rotary NP used as a micromirror to deflect light beam. The reflected beam rotates *2θ* when the nanoplate rotates *θ*. Distance between nanoplate and far field white screen is L (6.4 cm).

Relationship between position of laser spot on white screen (*y*) and rotation angle of reflected light (*2θ*) should be *y=L×tan(2θ)*. **b,** Sequencing optical images of the laser spot (the center of which is marked with red circles) on the screen in far field. **c,** Experimentally measured and theoretically expected position of laser spot on white screen. Rotational speed of the NP is 0.95 rpm (0.1 rad/s) actuated by light pulses at 5 kHz repetition rate in experiment. Preconceived relationship between y and t should be *y= L×tan(2ωt+θ$_0$)=6.4tan(0.2t+θ$_0$)*. *θ$_0$* is the initial angle.

We envision that the present nanoscale rotary locomotor provides unprecedented application potential in various fields, such as prospective micro-opto-electromechanical systems in outer-space, energy conversion, vacuum high-precision mechanics, etc. As already demonstrated, the rotating NP could serve as a scanning micromirror to deflect a laser beam, of which typical applications include laser scanning for miniature lidar systems[27] or laser display systems and optical modulating/switching for integrated micro-systems. Furthermore, the discovery of this novel light-actuated locomotion phenomenon could open a new era of optical driving and manipulation, with subnanometer locomotion resolution and controllable motion duration. It enables us to explore the new landscape of optical nano-manipulation in environments where those previous know-hows of liquid-based manipulation may not be translated.